\documentstyle[11pt]{article}

\def\eg{{\it e.g.}}

\begin{document}

% to type double-spaced full-width
\small
\renewcommand{\baselinestretch}{1.3}
\normalsize
%end redefinition

\title{Galaxy Candidates in the Zone of Avoidance}

\author{Ofer Lahav \thanks{Institute of Astronomy, Madingley Road, Cambridge CB3 0HA, UK} \and Noah Brosch$^{\dag}$   
\and Evgeny Goldberg \thanks{Department of Astronomy and Astrophysics and the Wise 
    Observatory, 
     School of Physics and Astronomy, 
    Raymond and Beverly Sackler Faculty of Exact Sciences, 
     Tel Aviv University, Tel Aviv 69978, Israel} \and George K.T. Hau$^*$ 
\and Ren\`{e}e C. Kraan-Korteweg \thanks{Observatoire de Paris-Meudon, 
D.A.E.C., 92195 Meudon Cedex, France} \and Andy J. Loan$^*$}

%\date{13 July  1997 \\ File zoa6.tex}
\date{15 July  1997 }
\maketitle

%\newpage

\begin{abstract}
Motivated by recent discoveries of nearby galaxies
in the Zone of Avoidance, 
we conducted a pilot study of galaxy candidates  
at low Galactic latitude, near Galactic longitude $l \sim 135^0$, 
where the Supergalactic Plane is crossed by the Galactic Plane. 
We observed with the 1m Wise Observatory 
in the I-band 
18 of the `promising' 
candidates identified by visual examination of Palomar red plates
by  Hau et al. (1995). 
A few candidates were also observed  in R or B bands,
or had spectroscopic
observations performed at the Isaac Newton Telescope
and at the Wise Observatory.
Our study suggests that there are probably 10
galaxies in this sample. We also identify a probable Planetary
Nebula.
The final confirmation of the nature of these 
sources must await the availability of full spectroscopic information.
The success rate of $\sim 50\%$  in identifying galaxies at Galactic latitude
$|b|<5^\circ$ indicates that the 
ZOA is a bountiful region to discover new galaxies.

\end{abstract}

\vspace{20mm}


\newpage

\section{Introduction}
Since the pioneering study of Kerr and Henning (1987) in searching for hidden
galaxies behind the Milky Way, a number of studies have been completed where
such objects have indeed been detected,  
\eg,  the Sagittarius dwarf (Ibata, Irwin and Gilmore 1994) and  
Dwingeloo 1 (Kraan Korteweg et al.  1994, Loan et al.  1996, 
Burton et al. 1996). 
%and the Antila dwarf (Whiting, Irwin and Hau  (1997).
These studies relied on optical photographic plates, 
HI 21 cm surveys, infrared searches, or X-ray
surveys (for reviews see Kraan-Korteweg and Woudt 1994; Lahav 1994, and 
Weinberger et al. 1995).

This paper studies some of the objects selected by 
Hau et al. (1995; hereafter HFLL) as galaxy candidates
from a visual search of
twelve Palomar Sky Survey (POSS) red (E) plates. The region, which covers
$\sim444$ square degrees
of the sky, is centered on Cassiopeia at $l\simeq135^{\circ}$
and $|b|\leq20^{\circ}$. It has been  surveyed with the intention 
of identifying a 
connecting structure of the Supergalactic Plane across the Galactic Plane, 
near the region of the Perseus-Pisces supercluster. HFLL
identified 2575 galaxy candidates, 
%of which 462 have major axis diameters
%larger than 0'.8. 
one of them being the galaxy Dwingeloo 1, 
detected 
at 21cm (Kraan-Korteweg et al. 1994). 
Our primary motivation in conducting the imaging and spectroscopic follow-up 
of this study is to check if other  nearby `obscured'  galaxies  appear 
in the HFLL sample.
The other purpose of this study is to estimate the number of galaxies
among candidates selected by visual inspection of photographic plates
at low Galactic latitude $(|b| < 5^\circ)$.
Although no obvious nearby galaxy was found, we are convinced that at least 
half of the candidates we observed are genuine galaxies.

\section{Observations}

Large objects from  the survey  of HFLL 
were scanned with the RGO PDS
from the POSS red glass copies. 
We selected objects from this large galaxy candidate subset that 
`looked' like galaxies on
hard copies of the PDS scans. The subjective criterion was the degree of 
`softness' of the edges of the candidate object, when compared to the stars
in the image; this is the equivalent to the subjective visual selection of the
candidates by HFLL, but on a contrast level that can be set to show
optimally what we believe are galaxies. 
Our 18 selected  candidates are listed in Table 1. Note that 15 of them 
are at low 
Galactic latitude $(|b|<5^\circ)$, a region where both obscuration and 
confusion with Galactic objects are strong.

Most objects were imaged during four nights in December 
1994 at the Wise Observatory (WiseObs); the images were flat-fielded and 
sky-subtracted at the site. The observations were performed with the
WiseObs Tektronix 1024$\times$1024 pixel chip, 
with  24 $\mu$m pixels, corresponding to 0".73 at the 
$f/7$ focus. 
Most observations were done in the I-band, as this spectral band 
samples the continuum in galaxies with $z \le 0.2 $ 
as well as in local emission nebulae. 
We expect 
galaxies to be bright in the I band because of the strong contribution from
late-type stars combined with the Galactic extinction. On the other hand,
emission nebulae will usually have only faint continua in the I band, because their
light is dominated by emission lines and is mostly in the R-band (H$\alpha$,
[NII] and [SII]), as well as in the V-band ([OIII] and H$\beta$). The same is
true of reflection nebulae, expected to be abundant close to the 
Galactic Plane. 
These would be much fainter in I than in bluer bands. Thus, while 
galaxies are expected to be bright both on the POSS plates (mainly on the E-plate)
and on our I-band images, emission or reflection nebulae will be hardly visible in our 
images despite being detected on the POSS images.
There is one caveat to this explanation; 
a reddened reflection nebula 
will appear on both images, but the likelihood of this happening and
still have a fairly smooth and symmetrical image is small.

A few objects were observed with the INT CCD camera during service nights in 
December 1994. These observations, as those from WiseObs, were in the I band.  
W. Saunders, O. Keeble, R. McMahon  and S. Maddox (private communication) 
kindly provided spectra of a few sources during an INT observing run
and a few spectra were also obtained with the WiseObs FOSC 
(Brosch \& Goldberg 1994).
Such observations are very useful in disentangling galaxies from emission
nebulae.

All our candidates were observed with the Dwingeloo 25 meter 
radio telescope, in the
same 21 cm band and velocity range as originally used 
for the Dwingeloo Obscured Galaxies Survey (Henning et al. 1997) i.e.
0-4000 km/sec  in 1 hour scans ($12 \times 5$ minutes 
in ON-OFF strategy), with flux limit of 40 mJy 
(rms per channel). 
In no case was a meaningful signal detected.
We note that a non-detection in 21cm does not necessarily imply that
an object is not a galaxy. It may have  low HI content or 
be at recession velocity larger than 4000 km/sec.

%OL: 
%[We also searched the ``All-Sky-Survey'' at 21 cm line from Dwingeloo (Butler Burton
%and ?) at the location of the candidates. This provides a cross-check of the low velocity
%range in HI, as well as extending our coverage to slightly negative velocities]
%NO DETECTION AMONG XX CANDIDATES, TO A FLUX LIMIT OF ?? FOR THE 
%VELOCITY RANGE UP TO 4000 KM/SEC.
%** TO RECHECK **

We also consulted catalogues for possible counterparts at different 
wavelengths,
 specifically searching for IRAS point sources which could
discriminate between Galactic dust and extragalactic sources.
 Any sources that emit at 12.5 and 
25$\mu$m but not at 60 and 100$\mu$m are {\it not} galaxies. 
Hau et al. (1994)
remarked on the FIR color segregation of sources 
that if $\frac{I_{100}}{I_{60}}>4$,
a source is most probably a Galactic cirrus cloud. Sources with 
$\frac{I_{60}}{I_{25}}<1$
and $\frac{I_{60}}{I_{12.5}}<1$ are likely to be  stars (Meurs \& Harmon 1988).

\section{Results}

\subsection{Individual sources}

    In this section we discuss the individual sources, based on the PDS
    scans of the POSS plates and on the images collected at WiseObs. 
Wherever we have spectroscopic information, we describe this
along with the appearance on the images. The PDS images of the objects are shown in Figures 1 and 2 and the WiseObs linear intensity 
images are shown in Figures 3-5. 
Table 1 gives
Equatorial and Galactic  
coordinates of the objects, 
and our proposed classification.
The discussion of individual objects includes also catalog information retrieved from
data banks (NED) within a 3 arcmin radius of each  candidate.
In the following, names of the objects are specified by the HFLL label 
(e.g., C36), as well as by Galactic coordinates (Glll.ll+bb.bb).

%ORDER LIST AND TABLE 1 BY RA

\bigskip
{\bf C36 (G132.45-3.38)}

This object was not included in our imaging sample because of the lack of
time at the telescope, but its spectrum 
 was  fortunately taken with the INT. It has
been classified by HFLL as an `elliptical' and identified 
as MCG 10-04-001, a galaxy with no given redshift,  0.1 arcmin away.
 Interestingly, it is the
{\it only} galaxy listed by Vorontsov-Velyaminov \& Krasnogorskaja (1962, MCG)
on this PSS plate. They call it a 17th mag, 1'.5 diameter 
overexposed 
image, with traces of structure which imitates the large-nucleus variety galaxy.  
Weinberger et al. (1995) quote a 0'.4 diameter of the red image.
%There is a radio source (87GB 020+5750), 4.5 arcmin away.

The INT spectrum shows
absorption features at $\lambda\lambda$6670, 5955, and a wide feature
at $\sim$5255\AA\,, identifiable   as
H$\alpha$, NaI 5890 and the Mgb band at a redshift of $\sim$0.015. The
spectrum has not been flux-calibrated, but has been corrected for atmospheric 
extinction and is strongly inclined  towards the red. 
The object is probably
a reddened galaxy at z$\approx$0.015, either the bulge of an early-type spiral or an 
E+A type, based on the spectroscopic evidence.

\bigskip
{\bf C39D1 (G133.63-3.62)} 

    Classified by HFLL  as dwarf elliptical, spiral or noise.
%    There is no NED identification.
    This is our best case of a clear-cut galaxy among the sources imaged
    in this pilot program. The PDS scan and our I-band image are virtually
    identical, showing a diffuse, round nebulosity centered on a bright,
    star-like object. The PDS scan shows a nearly round nebulosity; the
    I-band image is not fully symmetrical, but its Western side is brighter
    and more extended than the Eastern side. The I-band size of the
    nebulosity is 40"$\times$20". It is not clear why this object was
not selected as a candidate galaxy by Weinberger et al. (1995).

Unfortunately the INT spectrum is of poor quality.
This object warrants additional spectroscopic observations
to establish the identification as a galaxy and to measure its redshift.
%The INT spectrum shows a number of features similar to emission lines on
%a faint featureless continuum (COSMIC RAYS ???). 
%It is possible to identify two strong lines as
%H$\alpha$ and H$\beta$ at z$\approx$0.04, but three other lines,
%equally strong, 
%do not match any reasonable identification, precluding the association with
%any definite redshift. 
%There is no NED identification.

\bigskip
   {\bf C41 (G134.09-0.98)} 

    Classified by HFLL as dwarf elliptical or noise.
    The PDS scan showed at this location a very large, faint surface
    brightness patch, approximately centered on a group of three stars. The
    central group is definitely present on both the I and the B images we
    have for this object, but the diffuse patch is absent on both.
    It is possible
    that the object is a large, diffuse emission nebulosity, about 2 
    arcmin in size, which does not register in our I band. The lack of
    appearance on the B image is confirmed by the absence of the object on
    the O plate of the POSS. This is a case where either R-band or H$\alpha$
    imagery is called for. We can, however, conclude that this is not a good
    candidate for a galaxy. 
    Note that this is not considered 
    a galaxy candidate by Weinberger et al. (1995).

%Near this location there are two catalogued objects: BD+59 494 (2000) 02:21:38.6 
%+59:56:47.2 a V=9.3 star, and a (LBN) nebula at 02:21:39.5 +59:55:41.4. The
%location is close to an IRAS PSC source: 02171+5943 with marginal
%detection at 60 $\mu$m and $\frac{100}{60}>$4, indicative of cirrus emission.
%There is no NED identification.

\bigskip
  {\bf D30H15 (G135.74-4.53)}
 
    Classified by HFLL as elliptical or noise.
    The PDS scan is similar to the Wise I-band image,
    which shows a source elongated in the North-West to
    South-East direction. There is a possible slight 
    bending and a segment of a circle or ellipse. The appearance on
    the PDS scan of the E-plate of the POSS is of a diffuse patch,
    corresponding more or less to the brighter part of the circular segment
    in the I image. The source is also visible on the O plate. Based on
    this, we would classify it as a good candidate for a galaxy. 

%    Note also that this is a radio source in the Green Bank survey
%    (87GB 022051.9+554956), 3.3. arcmin away.
%About 3'E of the scanned candidate there is
%the 9.9 mag F8 star HD 14782, which is well separated from the source and 
%probably not associated with it.

\bigskip
   {\bf C9 (G133.87+2.53)} 
 
   This source appears extended on the POSS plates, but our I and B images
    indicate that the intensity profile is not different from that of other
    stars in the CCD field. The source consists of two stars, the northern
    one fainter in both bands than the southern. 
    Both the INT and WiseObs
    spectra indicate that
    both stars have Balmer absorption lines, with the fainter star having a
    comparatively bluer continuum. Based on the presence of strong H$\alpha$
    to H$\delta$ lines in the spectrum of the brighter component, we
    tentatively classify it as a late-B or early-A star. The H$\alpha$
    detection is confirmed by an INT spectrum, which
    also shows a NaI absorption line, both at rest velocity.
    The fainter component must, in this case,  be a late-O or early-B star. 
    The C9 candidate can fairly safely be identified as a pair of early-type
    stars (although classified by HFLL as 'elliptical').
    There is an IRAS PSC source 02272+6302 0.2 arcmin away from C9. 
% 1$^s$.9E and 9".1N of the C9. 
% fluxes are [60]=6.74 and [100]=27.9 Jy. 
Its FIR colour, $I_{100}/I_{60} \approx 4.0$,
  probably indicates some 
cirrus or cold dust contamination in the direction of these early-type stars.

\bigskip
    {\bf C26 (G135.91-0.47)}
 
    HFLL classified it as an `elliptical'. 
    The PDS scan of this source shows a $\sim$40" nebulosity centered
    on a bright star-like source. Our I band image shows no trace of the
    nebulosity, although the star-like source is present. In our view, and
    considering the difference in spectral bands between our I-band and the
    POSS E plate, this is an indication of an 
    emission nebula, either a planetary nebula (PN), or an HII region. These
    objects have almost no continuum emission, but only line contribution,
    where the strongest lines in the yellow-red region are [OIII],
    H$\alpha$, [NII] and [SII]. None will appear in our I band image, thus
    no nebulosity will be visible. We conclude that, for this source, our
    best interpretation is that it is some sort of Galactic emission nebula.
%    There is no NED identification.

%     NOT CONVINCING (LIKE C14J14, C8 ???)  

\bigskip
    {\bf C14J14 (G136.21+1.08)} 

    The appearance of this source on the PDS scan (also in Fig 13 of HFLL)
    is of a diffuse, almost
    circular patch with a bright centre.
    HFLL classified it as `unknown/elliptical' (but suggest a spiral 
    based on the PDS scan).
%    In I-band the object shows 
%   an elliptical
%    ring-like diffuse nebulosity, which is not connected with the bright
%    star at the West end of the diffuse ellipse. 
    The difference between the 	I-band image
    and the PDS scan may indicate a real difference in the source
    aspect when viewed in different spectral bands. An INT spectrum shows 
    that the source has strong H$\alpha$ absorption at rest velocity and a much
    weaker NaI, thus it can be safely identified as a star, perhaps A-type; 
    the diffuse source could be connected with this star.
%    There is no NED identification. 

\bigskip
   {\bf E94K86 (G141.06-9.30)} 

    Classified by HFLL as SB.
    The central object of  the I band image certainly confirms the appearance
    of the source on the PDS scan (Figures 1 and 2), also 
    shown in HFLL). The bright, small
    (40"$\times$20") nebulosity is only the central region of a larger
    spiral-like nebulosity, whose extension to the South and East can be
    discerned in the I image. The source coincides with UGC 2209,
    a 16.5 mag spiral galaxy 1.7 arcmin away with diameters 1'.0$\times$0'.6.
    Its diameter on the I image (1'.4)  is larger than 
    listed in the UGC. Our underexposed B image does not show the nebulosity.
    An INT spectrum shows Na I in absorption
    at $\lambda\simeq6055\pm10$\AA\, and a wide feature at $\sim$5320\AA\,, 
    identifiable as H$\beta$ and Mgb, 
    and indicating a redshift of $\sim$0.028.
    If indeed at this redshift, its I-band diameter corresponds to 
    $\sim 34$ kpc, suggesting a  big galaxy.

\bigskip
{\bf K88 and K89 (G141.08-9.22)}

 Classified by HFLL as 2 ellipticals.
  Our I-band image of E94K86 shows two other  diffuse images which
also show up in our shallow B-band image. In Figure 4
    the objects can be seen at  coordinates $(x \sim 0, y \sim 350$),
  as two diffuse `blobs', each about 30" in diameter. These look very
    much like a pair of elliptical galaxies and are listed in HFLL as 
    K88 and K89. We therefore conclude that these are also ``real'' galaxies.

HFLL proposed that E94K86 may have collided with the K88+K89 system;
in absence of redshift information on the binary system, and because of lack 
of evidence for K88+K89 being disrupted in any way, we cannot comment on this
possibility. In addition, if we accept the large extent of UGC 2209 as
indicated by our I-band image and its undisturbed appearance, there does
not appear to be any link between the objects. 
K88+K89 could well be its 
background. Also, close inspection of the PDS scan and of the
I-band image reveals
a number of faint, diffuse images in the vicinity of K88+K89. This is probably
evidence for a distant cluster of galaxies at this location,
and the pair K88+K89 could be its brightest central galaxies.

\bigskip
{\bf C8 (G135.62+2.76)}

This source was given a morphology of 
`star or unknown' by HFLL.
This source is extended on the POSS plates; it was not observed at 
the WiseObs, but a spectrum was secured at the 
INT. It shows clear signs of a Galactic emission nebulosity, with H$\alpha$, 
[NII] and [SII] emission at rest velocity. There is a strong somewhat
reddish continuum and a hint of H$\beta$ in absorption at the blue
end of the spectrum; we classify the object as an early-type star
with a compact HII region.
IRAS source 02421+6233 with $I_{100}/I_{60} \approx 1.4$ 
is 0.3 arcmin away.

\bigskip
  {\bf C11J7 (G135.80+2.72)}
 
   HFLL classified it as unknown/spiral. 
    This source appears as an East-West elongated diffuse streak in the PDS
    scan. We confirm this aspect from the I image,
    where the streak is at least
    two minutes long. The PDS scan (also shown in Fig. 13 in HFLL) indicates
    the 
    presence of a bright, star-like
    condensation close to the brighter part of the diffuse streak. It is
    possible to interpret this as the bulge of a disk galaxy close to edge-on. 
    In this case,
    and considering the bulge-to-disk ratio, the galaxy cannot be earlier
    than Sb. 
    We note that under heavy obscuration the disk shrinks more than the 
    bulge because of surface brightness effects. 
%    and remarked on the nearness of this object to Maffei I
%    ($\simeq3^{\circ}$.3 EARLIER VERSION???) 
     The object may be a member of the Maffei-IC342 
    group. The INT spectrum is underexposed and, unfortunately, does
    not allow a classification.
%    There is no NED identification.

%NO INTERESTING SOURCE IN W3 BROWSER

\bigskip
   {\bf J17 (G138.97+2.65)}

    Classified by HFLL as `unknown/SB'. 
    The PDS scan shows here a source with two condensations, that to
    North-West brighter than the second. A similar aspect is visible on the
    O-plate. Our I-band image confirms this description. The outer
    boundaries of the I-band image are about 2'$\times$40''. The
    location of the bright condensation is not central to the nebulosity; if
    it is not a superposed star, the more likely alternative is that we are
    viewing a system of interacting galaxies. The object is similar to Arp
    267, or 287, or 309. HFLL remarked that the proximity of this objects to Dw1 
    ($<3^{\circ}$) suggesting it may be part of the Maffei-IC342 group. The ZCAT
    lists for this object a velocity of 2350 km/sec 
(Huchtmeier \& Richter 1989),
which would argue for
    the object being in the background of the Maffei-IC342 group.
The IRAS PSC shows a nearby source (03067+6055),
% with fluxes[60]=2.82, 
%and [100]=6.51;
and the 60-to-100 $\mu$m ratio indicates this is a galaxy.
Radio source 87GB 030703.3+605537 is 2.4 arcmin away.

\bigskip
 {\bf J4 (G139.31+4.84) and J5 (G139.32+4.85) }
 
    We combine these two sources not only because they appear on the same I-band
    image, but also because we believe them to be parts of the same emission
    nebulosity. The sources appear as two round condensations on the PDS
    scan, located nearly symmetrically around a brighter star. There is a
    faint common envelope around both sources, which is $\sim$90"$\times$35".
    The I and R images are shown in Figure 5.
    Our R-band image is almost identical with this description, but
    the I-band image, although with stretched contrast, does not show the
    source. This is another case where we believe that the sources are parts
    of an emission nebula, perhaps a two-lobe planetary nebula (PN). 

The IRAS PSC lists a source reasonably nearby (03188+6236) with [100]=11.79
and marginal detection at 60 $\mu$m; 
this indicates again the presence of cold dust near the
candidate, thus a  Galactic origin. Lacking a spectrum of the star we cannot
be certain of the nature of the nebulosity; however we tend to believe this
to be a small HII region, or a PN. 
% Radio source 87GB 0318+6237 is 2.4 arcmin away.
If its size is  typical for PNs ($\sim 1 $pc ),
 its distance could be $\sim 2.3$ kpc.

\bigskip
   {\bf J22 (G140.72+3.02) }

     Classified by HFLL as noise or dwarf elliptical. 
    This is an elongated patch on the PDS scan, which is visible also on the
    O plate, but is not visible on our I-band image. Our interpretation is
    again of an emission nebula, which has no continuum in the I-band. 
%    No
%    IRAS, radio, or X-ray sources were found in the neighborhood.

\bigskip
    {\bf J18 (G140.93+3.98) and J19 (G140.98+4.02) }
 
   Classified by HFLL as 'unknown' and 'unknown/spiral` respectively.  
    The two objects appear on the same I-band images and both are good
    candidates for galaxies. The size of J18 is $\sim$90"$\times$40"
    and the major axis PA is $\sim30^{\circ}$. J19 is smaller, 
   only $40''\times 20''$, 
   and its major axis is at PA$\simeq 0^\circ$. We believe both to be disk
    galaxies, perhaps Sa or S0. 

The INT spectrum is underexposed and shows no features useful for classification.
The radio source 87GB 032605.9+605943 is 1 arcmin away and
is possibly
connected with the two galaxies. The IRAS PSC contains a source 
IRAS03264+6100 close to the pair. 
Its FIR colour 
indicates cirrus emission.

\section{Discussion}

 We observed 18 candidate objects from HFLL, selected on 
basis of their appearance on red images, and classified  with 
reasonable degree of confidence 10 as possible galaxies.  
Four objects are probably stellar, within the Milky Way,  one is a probable
planetary nebula, and 2 are of uncertain nature. The $\sim  50\%$ 
 success rate in finding objects at Galactic latitude 
$|b|<5^\circ$ which ``look'' like
galaxies, some of which were confirmed as such by spectra, indicate that the
Zone of Avoidance (ZOA) is not completely devoid of clear patches through which
one can look out of the Milky Way.
We note that in other searches the success rate of classifying
galaxies among visually selected candidates 
in the ZOA was reported to be higher  
(e.g. 98 \%, Kraan-Korteweg et al. 1994). 
 
In retrospect, 
we recognize the importance of the I-band as the primary determinant
of the nature of a source. 
The images in this band sample {\bf only} the continuum
both for $z<0.2$ galaxies and for Galactic nebulae, 
whereas the POSS E-plates sample not only
continuum but also emission line contributions. Combining the information about
the aspect of a source on the POSS E-plate and in our I-band images, 
we were able to
recognize which sources were galaxies and which were emission nebulae.

For future studies, it may be useful to add an H$\alpha$ filter at rest
wavelength in place of the R-band used here sometimes. A comparison with the
I image would show even better
which sources are emission line objects. 
The proposed 
$H\alpha$ survey of the Galactic Plane 
with the UK Schmidt telescope (Parker et al. 1995) 
can be used to reject HII regions and 
Planetary nebulae.

The importance of the search for 
galaxies and clusters of galaxies in this direction lies in that it samples
a region along the Supergalactic Plane, near the Perseus-Pisces supercluster  
in the opposite direction of the Great Attractor.  It will therefore be
interesting to probe the space distribution of galaxies there.

\section*{Acknowledgements}
We thank
J. Pilkington for the PDS scans,
 W. Saunders, O. Keebles, R. McMahon and  S. Maddox for obtaining the INT spectra, 
and  P. Henning and B. Burton for their part in the Dwingeloo observations.
Observations at the Wise Observatory are supported by a Center of Excellence
Award from the Israel Science Foundation.
OL acknowledges the hospitality of the Wise Observatory.

\section*{References}

\begin{description}

\item Brosch, N. \& Goldberg, E. 1994, MNRAS, 268, L27

\item Burton, W.B., Verheijen, M.A.W., Kraan-Korteweg, R.C., 
 Henning, P.A., 1996, A\&A, 309, 687

\item Hau, G.K.T., Ferguson, H.C., Lahav, O. \& Lynden-Bell, D. 1995, 
MNRAS,  277, 125

\item Henning, P.A., Kraan-Korteweg, R.C., Rivers, A.J., 
Loan, A.J., Lahav, O., Burton, W.B.,  1997, in preparation 

\item Huchtmeier, W.K. \& Richter, O.G., 1989, 
``A General Catalogue of HI observations of Galaxies'', New York, 
Springer-Verlag 

\item Ibata, R., Irwin, M., \& Gilmore, G. 1994, Nature, 370, 194

\item Kerr, F. \& Henning, P.A. 1987, ApJ, 320, L99

\item Kraan-Korteweg, R.C., Loan, A.J., Burton, W.B., Lahav, O., 
Ferguson, H.C., Henning, P.A. \& Lynden-Bell, D. 1994, Nature  372, 77

\item Kraan-Korteweg, R.C. \& Woudt, P.A., 1994, 
in {\it Unveiling large-scale
structures behind the Milky Way} (C. Balkowski \& R.C. Kraan-Korteweg, eds.),
A.S.P. Conference Series, p. 89

\item Kraan-Korteweg, R.C., 1994,
in {\it Unveiling large-scale
structures behind the Milky Way} (C. Balkowski \& R.C. Kraan-Korteweg, eds.),
A.S.P. Conference Series, p. 99 
 
\item Lahav, O. 1994, in {\it Unveiling large-scale
structures behind the Milky Way} (C. Balkowski and R.C. Kraan-Korteweg, eds.),
A.S.P. Conference Series, p. 7

\item Loan, A.J., Maddox, S.J., Lahav, O., Balcells, M., Kraan-Korteweg, R.C., 
Assendorp, R., Almoznino, E., Brosch, N., Goldberg, E. \& Ofek, O., 1996, 
MNRAS, 283, 537

\item Meurs, E.J.A. \& Harmon, R.T. 1988 A\&A,  206, 53

\item Parker, Q.A., Phillips, S., Morgan, D.H., 1995, 
IAU colloquium No. 148, ASP Conference Series 84, J.M. Chapman et al., eds., 
p. 129.

\item Vorontsov-Velyaminov, B. \& Krasnogorskaja, A. 1962 Trans. State Astron. Inst
Shternberg Vol. XXXII, MCG=Morphological Catalogue of Galaxies.

\item Weinberger, R., Saurer, W. \& Seeberg, R. 1995, 
 A\&AS, 110, 269

%\item Whiting, A., M. Irwin \& Hau,G. 1997, AJ, in press

\end{description}

\newpage

\section*{Table 1: Source classification}
\begin{tabular}{|l|l|l|l|l|l|l|}
\hline
Object & $\alpha$(1950) & $\delta$(1950) & l & b & Observations & Proposed classification \\ \hline
C36    & 02:00:50.6 & 57:54:06 & 132.4 & -3.4 & Is & Galaxy \\
C39D1  & 02:08:46.7 & 57:19:50 & 133.6 & -3.6 & Wi, Is & Galaxy \\
C41    & 02:18:35.7 & 59:41:11 & 134.1 & -1.0 & Wi & No ID:  plate defect ? \\
D30H15 & 02:21:02.4 & 55:47:02 & 135.7 & -4.5 & Wi & Galaxy ?\\
C9     & 02:27:14.2 & 63:02:32 & 133.9 & +2.5 & Wi, Ws, Is & Two stars and nebulosity \\
C26    & 02:33:25.8 & 59:29:34 & 135.9 & -0.5 & Wi & Star+ emission nebula \\
C14J14 & 02:40:41.2 & 60:47:07 & 136.2 & +1.1 & Wi, Is & Star \\
E94K86 & 02:41:02.5 & 49:20:22 & 141.1 & -9.3 & Wi, Is & Galaxy \\
K88    & 02:41:18.7 & 49:24:27 & 141.1 & -9.2 & Wi & Galaxy, cluster ? \\ 
K89    & 02:41:18.7 & 49:24:27 & 141.1 & -9.2 & Wi & Galaxy, cluster ? \\
C8     & 02:42:06.7 & 62:33:45 & 135.6 & +2.8 & Is & Star+HII \\
C11J7  & 02:43:22.6 & 62:27:06 & 135.8 & +2.7 & Wi, Is & Galaxy ?\\
J17    & 03:06:43.6 & 60:55:17 & 139.0 & +2.7 & Wi & Galaxy/interacting pair \\
J4+J5  & 03:18:55.7 & 62:36:41 & 139.3 & +4.8 & Wi & Planetary nebula/HII region\\
J22    & 03:20:24.9 & 60:18:56 & 140.7 & +3.0 & Wi & No ID \\
J18    & 03:26:13.9 & 60:59:59 & 141.0 & +4.0 & Wi, Is & Galaxy ? \\
J19    & 03:26:41.8 & 61:00:07 & 141.0 & +4.0 & Wi, Is & Galaxy ? \\
\hline
\end{tabular}

\vspace{10mm}

Notes to Table 1:
\begin{itemize}
\item Wi=WiseObs imaging (mostly in the I band)
\item Ws=WiseObs spectra
% \item Ii=INT imaging
\item Is=INT spectra
\end{itemize}

\end{document}